\documentstyle[twocolumn,prb,aps,epsfig]{revtex}
\begin{document}

\ifpreprintsty\else
\twocolumn[\hsize\textwidth%
\columnwidth\hsize\csname@twocolumnfalse\endcsname
\fi

\title{Electrical transport through carbon nanotube junctions created by mechanical manipulation} 

\author{Henk W.Ch. Postma, Mark de Jonge, Zhen Yao, and Cees Dekker} 

\address{Department of Applied Physics and DIMES, Delft University of Technology, 
Lorentzweg 1, \mbox{2628 CJ Delft}, The Netherlands}

\maketitle

\begin{abstract}
Using an atomic force microscope we have created nanotube junctions 
such as buckles and crossings within individual single-wall metallic carbon 
nanotubes connected to metallic electrodes. The electronic transport properties of 
these manipulated structures show that they form electronic tunnel junctions. 
The conductance shows power-law behavior as a function of bias voltage and 
temperature, which can be well modeled by a Luttinger liquid model for tunneling 
between two nanotube segments separated by the manipulated junction. 
\end{abstract}

\ifpreprintsty\clearpage\else\vskip1pc]\fi
\narrowtext

\markright{H.W.Ch. Postma {\em et al.}}

Molecular electronics has taken a large step forward since the discovery of 
carbon-nanotube metallic and semiconducting molecular wires \cite{cees}. Various 
nanotube devices have been found to behave as conventional electronic components. 
For instance, individual semiconducting nanotubes function as field-effect 
transistors at room temperature \cite{tansfet}, while metallic nanotubes are 
single-electron transistors at low temperature \cite{bockrathropes,tansspectr}. More 
recently, it was found that intramolecular metal-semiconductor kink junctions can 
act as rectifying diodes at room temperature \cite{yao}. Unlike conventional 
solid-state devices, however, nanotubes are molecules. Conformational changes can 
therefore be expected to strongly affect the electronic properties of nanotubes, 
opening up a route towards nanoscale electro-mechanical devices (NEMs). Indeed, theoretical work has indicated that local deformations such as twists and buckles may induce strong barriers for electron transport \cite{rochefort,kanemele,bernholcprb}. While some transport experiments have been conducted on carbon nanotube junctions which occur naturally\cite{yao,lefebvre,fuhrer} and on defects due to locally applied strain\cite{daiinsitu}, a focussed study with control over the geometry and configuration of the junction is lacking. 

Here, we report electron transport measurements on molecular junctions that have 
been fabricated in a controlled manner from straight undeformed nanotubes by 
manipulation with an atomic force microscope (AFM). We have fabricated nanotube 
buckles and crossings and characterized their electron transport properties. We 
find that these mechanically manipulated structures act as tunnel junctions with a 
conductance that show power-law dependences on both bias voltage and temperature. 
For various sample layouts we obtain a wide range of power-law exponents, from 0.25 
to 1.4. We show that this variety can be understood within one consistent 
Luttinger model.

Single-wall carbon nanotubes were produced by the group of R.E. Smalley at Rice University, USA. A small amount of this raw material is ultrasonically dispersed and spin coated on top of a SiO$_2$/Si-substrate containing a large array of predefined Pt electrodes. These electrodes are fabricated using a  double layer polymethylmethacrylate/metacrylic acid (PMMA/MAA) resist, electron beam lithography, reactive ion etching,  Pt evaporation and lift-off. The resulting electrodes are embedded in the SiO$_2$ substrate such that the height difference between the electrodes and substrate is less than 1 nm. Nanoscale tunnel junctions are then created within individual carbon nanotubes by use of the AFM. Conductance measurements are performed using a standard ac-lockin technique. 
                                       
\begin{figure}[h]
\centerline{\epsfig{file=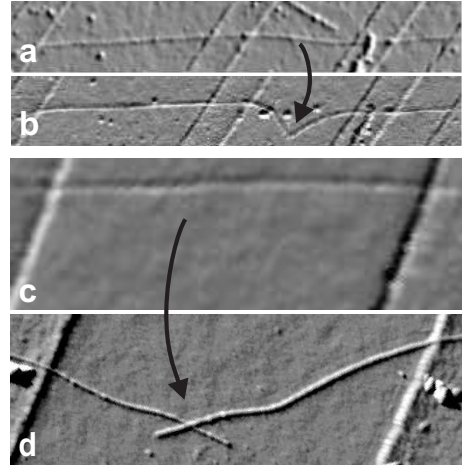, width=6cm, clip=}} 
\vspace*{0.4cm} 
\caption{ \label{fig1} Formation of carbon nanotube nanojunctions by AFM 
manipulation. Between the images in {\em a} and {\em b}, an initially straight 
nanotube has been dragged to the bottom by the AFM tip, resulting in a sharp 
105$^\circ$ buckle. Image {\em c} and {\em d} show the manipulation of a nanotube 
crossing from an initially straight nanotube. The nanotube ends are extending 110 
(left) and 130 nm (right) beyond the crossing point. The difference in apparent 
width of the nanotubes in these images is due to variation in the AFM tip radius 
which is different for different tips, and which moreover can change in the 
manipulation process.  } 
\end{figure} 

\begin{figure}[h]
\centerline{\epsfig{file=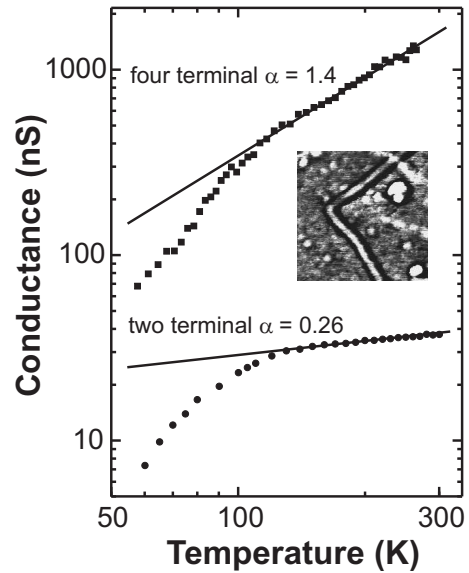, width=6cm, clip=}} 
\vspace*{0.4cm}
\caption{ \label{fig2} Conductance of a nanotube buckle as a function of 
temperature in a four- and two-terminal measurement. The straight solid lines on 
this log-log plot indicate the power-law behavior $G \propto T^\alpha$, with the 
exponent $\alpha$ as denoted. Below 120 K, Coulomb blockade sets in which further 
suppresses the conductance at low temperatures. The inset shows a 300$\times$300 
nm$^2$ AFM phase image of the nanotube buckle.  
The four-terminal measurement reveals the intrinsic buckle conductance, whereas the 
two-terminal conductance is limited by the contact conductance. } 
\end{figure}

Figure \ref{fig1} presents two examples of nanojunctions that were fabricated with 
an AFM from individual metallic carbon nanotubes. In the fabrication procedure, the 
tip of the AFM is used to change the lateral position of a nanotube lying on top of 
metallic electrodes. First, a nanotube is identified by scanning the tip over the 
sample in tapping-mode AFM. Then, the tip is pressed onto the surface and moved 
along a predefined path across the nanotube. In this manner, the position and shape 
of nanotubes can be controlled with a high degree of accuracy \cite{manipulation}. 
In Fig. \ref{fig1}a we show the initial configuration of a straight nanotube lying 
across four electrodes. In order to bend the tube between the middle two electrodes, 
the nanotube has been dragged across the surface in a direction perpendicular to its 
length. During this dragging action, the nanotube has slided along its length across 
the electrodes. The sharp bend that results from the AFM manipulation has an 
angle of 105$^\circ$ (see Fig.\ref{fig1}b, and also inset to Fig. \ref{fig2}). This 
is well above the critical value of about 60$^\circ$ needed to form a so-called 
`buckle' \cite{iijimatwo}, where a strongly bent nanotube releases strain by 
locally collapsing the cylindrical shell structure into a flattened tube structure. 
Accordingly, a small height increase is found at the bending point. Another example 
of a manipulated nanojunction is shown in Fig. \ref{fig1}c,d. In this case, the 
dragging action of the AFM has broken the nanotube. Subsequently, the two broken 
ends of this nanotube have been pushed back together into a configuration where they 
cross each other. The resulting nanotube ends extend about 100 nm beyond the 
crossing point. 

Multi-terminal contacting of the nanotube allows one to separately measure the 
contact conductance (from two- and three-terminal measurements) and the intrinsic 
conductance of the manipulated tube (from a four-terminal measurement). The buckled 
nanotube sample in Fig. \ref{fig1}b has contacts with a low contact conductance, 
i.e., only 65 nS at room temperature. The intrinsic buckle conductance appears to 
be about 1 $\mu$S at room temperature. This is much lower than the four-terminal 
conductance value of order 100 $\mu$S that we typically find for non-manipulated 
straight nanotubes in a similar layout. The effect of the buckle on the electron 
transport is thus quite dramatic. The buckle conductance is also much lower than the 
quantum conductance unit of $4e^2/h = 154$ $\mu$S, which indicates that the buckle 
acts as a tunnel barrier. 

In Fig.\ref{fig2} the conductance $G$ of the buckled segment is plotted versus 
absolute temperature $T$ on a double-logarithmic scale for both the two- and four-terminal 
configuration. At high temperatures the data can be fitted with a power-law function $G \propto 
T^\alpha$ (solid lines). Below \mbox{120 K}, Coulomb blockade sets in which further suppresses the conductance. The power-law exponent $\alpha$ is found to be very different, $\alpha=0.26$ versus $1.4$, for the two- and 
four-terminal measurements respectively. The intrinsic buckle conductance 
(four-terminal data) thus appears to be much more strongly temperature dependent 
than the contact conductance (two-terminal data).

We can understand these findings on the basis of a Luttinger liquid model. The 
Luttinger model \cite{kane,egger} has been employed to explain recent transport 
experiments on metallic carbon nanotubes \cite{luttinger,yao}. In this model, 
electron-electron correlations combined with the one-dimensional nature of nanotubes 
lead to a power-law suppression of the tunneling conductance as a function of 
energy, $dI/dV \propto E^\alpha$. Here $E$ is the maximum of the thermal or voltage 
energy scale, i.e. $k_B T$ or $eV$ respectively, with $k_B$ Boltzmann's constant and 
$e$ the electron charge. At low bias voltages  $V\ll k_BT/e$ this leads to a 
power-law behavior of the conductance as a function of $T$, i.e., $G \propto 
T^\alpha$. At high voltages $V\gg k_BT/e$, however, it yields a power-law dependence on voltage, 
$dI/dV \propto V^{\alpha}$. The exponent $\alpha$ depends on the strength of the 
electron-electron interactions which is characterized by the Luttinger interaction 
parameter $g$. \cite{kane,egger,luttinger} For repulsive interactions, $g$ ranges from 0 (very strong 
interactions) to 1 (no interactions). Estimates of $g$ for carbon nanotubes are in 
the range of 0.2 -- 0.3. \cite{kane,egger,luttinger} The exponent $\alpha$ also 
depends on the position of tunneling. When electrons are added to the end of the 
nanotube, the excess electron charge can spread away in one direction only and the 
tunnel conductance is suppressed strongly with an exponent $\alpha_{end} = 
(1/g-1)/4$. Tunneling into the bulk of the nanotube is more weakly suppressed, with 
$\alpha_{bulk} = (1/g + g-2)/8$, because the excess charge can now spread in both 
directions away from the contact.  

The conductance of the buckle is suppressed with a power-law exponent $\alpha=1.4$ 
(Fig. \ref{fig2}). If the buckle acts as a tunnel barrier, transport across the 
buckle takes place by tunneling of electrons from the end of one nanotube segment to 
the end of the other segment. This end-to-end tunneling is associated with an 
exponent twice as large as tunneling into a single end, i.e., $\alpha_{end-end} = 
2\alpha_{end} = (1/g-1)/2$. Solving $\alpha_{end-end}=1.4$ yields a Luttinger 
interaction parameter value $g=0.26$. In the {\em two}-terminal configuration, 
however, the contacts limit the conductance and one thus probes bulk tunneling from 
the contacts to the nanotube. Here we find $\alpha_{bulk} = 0.26$, from which we 
obtain the {\em same} Luttinger parameter value $g=0.26$. It is gratifying that 
these exponents which are differing by a factor 6, can be reconciled by this single 
parameter $g$. The value of $g=0.26$ is also well in agreement with theoretical 
estimates \cite{kane,egger}, recent experiments in a different geometry \cite{yao}, 
and the value of $g=0.29\pm  0.04$ that we find for many samples with straight 
non-manipulated nanotubes. We thus conclude that the transport characteristics of 
this buckle are well described by assuming that it acts as an artificially created 
nanometer-size tunnel junction within an individual nanotube.

\begin{figure}[h]
\centerline{\epsfig{file=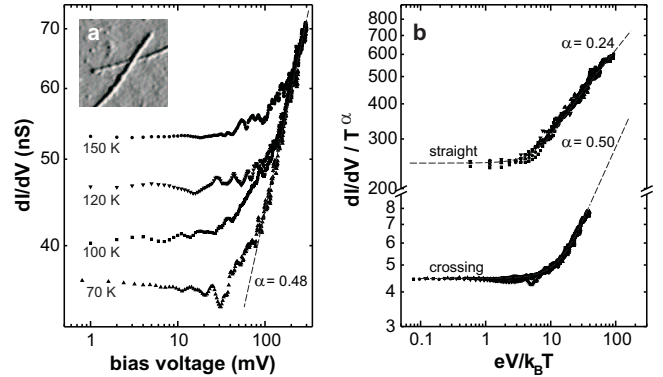, width=8.6cm, clip=}} 
\vspace*{0.4cm} 
\caption{ \label{fig3} Differential conductance of a manipulated nanotube crossing 
as a function of applied bias voltage for several temperatures. For this sample, the Coulomb blockade effect suppresses the conductance below 70 K. At low bias 
voltages, ${dI}/{dV}(V)$ is constant while it depends as a power-law on temperature. 
At high voltages, the differential conductance crosses over to a power-law 
dependence on bias voltage ${dI}/{dV}\propto V^\alpha$, with $\alpha=0.48$ (dashed 
line). The inset of {\em a} shows a 200$\times$200 nm$^2$ AFM amplitude image of the 
crossing. Figure {\em b} presents a scaling plot, where ${dI}/{dV}$ has been scaled by $T^\alpha$ and is plotted versus $eV/k_BT$ for the crossing segment and, for comparison, for a typical straight segment of a nanotube.  } 
\end{figure}

We now discuss data for the nanotube-crossing sample shown in Fig.1d. The 
conductance of the crossing reads \mbox{80 nS} at room temperature 
\cite{crossinghistory}. Again this value is much lower than the conductance quantum 
indicating that the crossing also acts as a tunnel junction. The conductance again 
decreases as a power-law upon lowering the temperature, with $\alpha=0.50$ (not 
shown). For this sample, the Coulomb blockade effect further suppresses the conductance below 70 K. The bias dependence of the differential conductance at several temperatures 
is shown in Fig. \ref{fig3}. At all temperatures, the data show the same behavior: 
At low applied bias, $dI/dV$ is constant at a level that scales as a power law with 
temperature ($\alpha=0.50$). At high bias voltage it crosses over to a power-law 
voltage dependence, i.e., $dI/dV \propto V^{\alpha}$ with $\alpha=0.48$ (dashed 
line). The dependence of the differential conductance on both energy scales $eV$ 
and $k_BT$ is emphasized in Fig. \ref{fig3}b, where the differential conductance is 
scaled by $T^\alpha$ and plotted versus $eV/k_B T$. As expected, all the data obtained at different temperatures  and bias voltages collapse onto a single curve, which is well described by the theoretically expected form (dashed line) \cite{scalingfunction}. The exponent $\alpha$ that has been used to scale these curves onto each other is 0.50. Transport between crossing nanotubes was studied recently, but only in the low-bias regime, where this power-law behavior was not observed \cite{fuhrer}. 

The crossing junction thus yields a significantly different value, $\alpha \approx 
0.50$, than the buckle junction discussed above. This can be understood as a direct 
consequence of the particular crossing geometry. Unlike the case for the nanotube 
buckle where the two tube ends meet, the contact in the crossing is now from the 
bulk of one tube to the bulk of the other. The electron transport thus takes place 
via bulk-to-bulk tunneling \cite{bulk} with an exponent that is twice as large as 
that for regular bulk tunneling, i.e., $\alpha_{bulk-bulk} = 2\alpha_{bulk} = 
(1/g+g-2)/4$. From $\alpha_{bulk-bulk}=0.50$ we find $g=0.27$, which again is in 
excellent agreement with the other results for $g$. The bulk-to-bulk tunneling 
observed for the crossing can be readily compared to the regular bulk-tunneling 
configuration, which is done in Fig.\ref{fig3}b, where the scaled differential 
conductance is shown  for a straight nanotube as well \cite{scalingfunction}. 
In this case, the exponent is found to be $\alpha = 0.24$, which indeed is half the exponent observed for bulk-to-bulk tunneling. Molecular dynamics simulations have suggested that crossing nanotubes can be both deformed by about \mbox{20\%} at the crossing point due to the van der Waals binding of the upper nanotube to the substrate away from the crossing \cite{herteltwo}. Apparently, this deformation, if present at all, does not electronically break up the nanotubes, since our data indicate that intertube transport occurs via bulk-to-bulk rather than through end-to-end or end-to-bulk tunneling. 

Recently, transport experiments were conducted on metal-metal nanotube kink 
junctions formed by a pentagon-heptagon defect pair located at the kink \cite{yao} and naturally occurring crossing junctions\cite{lefebvre,fuhrer}. 
Whereas such junctions are rare objects, the present work shows that one can 
use an AFM to precisely define local junctions at arbitrary positions along a 
nanotube. The transport characteristics demonstrate that these local junctions 
significantly alter the electronic transport properties of carbon nanotubes. A 
unifying description of single nanotubes, kinks, buckles, and crossings can be 
obtained from the Luttinger liquid model. The manipulation technique shown here 
allows the fabrication of various interesting new nanotube structures. For instance, 
double-buckle structures can be envisioned which define a room-temperature 
single-electron transistor \cite{chico,doublebuckle}. More generally, we expect 
that electro-mechanical effects may find their use in future nano-electronic 
devices.

We thank A. van den Enden for experimental assistance and L. Balents for useful
discussions. The nanotube material was kindly supplied by R.E. Smalley and
coworkers at Rice University, USA. This research is financially supported by the Dutch Foundation for Fundamental Research on Matter (FOM) and the European Community SATURN Project.


\begin{thebibliography}{10}

\bibitem{cees}
C.~Dekker,
\newblock { Physics Today} {\bf 52}, 22 (1999)

\bibitem{tansfet}
S.J. Tans, A.R.M. Verschueren, and C.~Dekker,
\newblock { Nature} {\bf 393}, 49 (1998)

\bibitem{bockrathropes}
M.~Bockrath {\it et al.},
\newblock { Science} {\bf 275}, 1922 (1997)

\bibitem{tansspectr}
S.J.~Tans {\em et al.},
\newblock { Nature} {\bf 386}, 474 (1997)

\bibitem{yao}
Z.~Yao, H.W.Ch. Postma, L.~Balents, and C.~Dekker,
\newblock { Nature} {\bf 402}, 273 (1999)

\bibitem{rochefort}
A.~Rochefort, D.R. Salahub, and Ph. Avouris,
\newblock { Chem. Phys. Lett.} {\bf 297}, 45 (1998)

\bibitem{kanemele}
C.L. Kane and E.J. Mele,
\newblock { Phys. Rev. Lett.} {\bf 78}, 1932 (1997)

\bibitem{bernholcprb}
M.B. Nardelli and J.~Bernholc,
\newblock { Phys. Rev. B} {\bf 60}, R16338 (1999)

\bibitem{lefebvre}
J.~Lefebvre {\em et al.},
\newblock { Appl. Phys. Lett.} {\bf 75}, 3014 (1999)

\bibitem{fuhrer}
M.S.~Fuhrer {\it et al.},
\newblock { Science} {\bf 288}, 494 (2000)

\bibitem{daiinsitu}
T.W.~Tombler {\it et al.},
\newblock { Nature} {\bf 405}, 769 (2000)

\bibitem{manipulation}
H.W.Ch. Postma, A.~Sellmeijer, and C.~Dekker,
\newblock { to appear in Adv. Mater.} (2000)

\bibitem{iijimatwo}
S.~Iijima, C.~Brabec, A.~Maiti, and J.~Bernholc,
\newblock { J. Chem. Phys.} {\bf 104}, 2089 (1996)

\bibitem{kane}
C.~Kane, L.~Balents, and M.P.A. Fisher,
\newblock { Phys. Rev. Lett.} {\bf 79}, 5086 (1997)

\bibitem{egger}
R.~Egger and A.O. Gogolin,
\newblock { Phys. Rev. Lett.} {\bf 79}, 5082 (1997)

\bibitem{luttinger}
M.~Bockrath {\em et al.},
\newblock { Nature} {\bf 397}, 598 (1999)

\bibitem{crossinghistory}
Initially the crossing nanotube structure was contacted by four electrodes,
  with the crossing located between the middle two electrodes. The conductance
  of each of the two middle electrodes were determined to be about 250 nS at
  room temperature. After the initial measurements, however, the outer contacts
  were lost, and two-terminal measurements are reported hereafter.

\bibitem{scalingfunction}
The scaling functions read $(dI/dV) / T^\alpha $ $ \propto $ \/ $\sinh(\eta
  x/2)\left| \Gamma\left(1+\alpha/2+i\eta x/2\pi\right)\right|^2 \times$
  $\left\{ \frac{1}{2} \coth(\eta x/2) - \frac{1}{\pi} \mbox{Im}
  \left[\Psi(1+\alpha/2+i\eta x/2\pi) \right] \right\} $ for the bulk-to-bulk
  tunneling expected for the crossing, and $dI/dV /T^\alpha \propto
  \cosh(\gamma x/2) \left| \Gamma\left( \frac{1+\alpha} {2}+i\gamma
  x/2\pi\right)\right|^2$ for a straight tube. Here, $x\equiv eV/k_BT$,
  $\gamma$ depends on the ratio of the contact conductances, $\eta$ accounts
  similarly for the fraction of the applied bias voltage that drops at the
  contacts to the nanotube, and $\Gamma$ and $\Psi$ are the gamma and digamma
  functions, respectively. Using $\gamma$ and $\eta$ as fitting parameters, we
  find $\gamma=0.6$ and $\eta=0.18$.

\bibitem{bulk}
A contact qualifies as a bulk contact if the distance from the contact to the
  end of the nanotube $L > \frac{h v_F}{2\pi k_BT}$, where $v_F$ is the
  Fermi-velocity and $h$ is Planck's constant \cite{orthogonal}. In the sample
  presented in Fig. 1d, the ends are separated from the crossing point by 110
  and 130 nm, which means that all measurements performed at temperatures above
  56 K will show bulk-to-bulk tunneling.

\bibitem{herteltwo}
T.~Hertel, R.E. Walkup, and Ph. Avouris,
\newblock { Phys. Rev. B} {\bf 58}, 13870 (1998)

\bibitem{chico}
L.~Chico, M.P.~L\'opez Sancho, and M.C. Munoz,
\newblock { Phys. Rev. Lett.} {\bf 81}, 1278 (1998)

\bibitem{doublebuckle}
M.S.C. Mazzoni and H.~Chacham,
\newblock { Phys. Rev. B} {\bf 61}, 7312 (2000)

\bibitem{orthogonal}
L. Balents, cond-mat/9906032. To appear in {\em XVIII Moriond Les Arcs
  Conference Proceedings}, edited by D. C. Glattli and M. Sanquer (Edition
  Frontiers, France, 1999)

\end{thebibliography}
\end{document}